\documentclass[prl,twocolumn,showpacs,floatfix]{revtex4}
\usepackage{amsmath}
\usepackage{dcolumn}
\usepackage{graphicx}
\usepackage{natbib}

\newcommand{\ket}         [1] {\big|#1\big\rangle}
\newcommand{\bra}         [1] {\big\langle#1\big|}

\newcommand{\average}     [1] {\big\langle#1\big\rangle}

\newcommand{\eee} {\mathcal{E}}
\newcommand{\eeee}[2]{\eee_{#1} (\omega_{#2} - \omega_{#1})}

\begin{document}

\title{Coherent ultrafast core-hole correlation spectroscopy; x-ray
analogues of multidimensional NMR}
\author{Igor V.~Schweigert}
\author{Shaul Mukamel}
\affiliation{Department of Chemistry, University of California, Irvine, California
92697-2025}

\begin{abstract}
We propose two dimensional x-ray coherent correlation spectroscopy (2DXCS)
for the study of interactions between core-electron and valence transitions.
 This technique might find experimental applications in the future when
very high intensity x-ray sources become available.  Spectra obtained by
varying two delay periods between pulses show off-diagonal cross-peaks
induced by coupling of core transitions of two different types. Calculations
of the N1s and O1s signals of aminophenol isomers illustrate how novel
information about many-body effects in electronic structure and excitations
of molecules can be extracted from these spectra.
\end{abstract}

\pacs{33.20.Rm, 42.65.Re}
\maketitle

X-ray absorption spectroscopy (XANES, EXAFS)\cite{Stohr1996} and its
time-resolved
extensions\cite{ChenJagerJenningsEtAl2001,BresslerChergui2004,ChenZhangTomovEtAl2007}
provide a direct probe for electronic structure of molecules with
subatomic and subfemtosecond resolution. Novel ultrabright x-ray
sources such as future free-electron laser (XFEL) \cite{LCLS} or
high-harmonic (HH) sources \cite{KapteynCohenChristovEtAl2007} may
make it possible to perform nonlinear experiments with multiple x-ray
pulses. All-x-ray nonlinear signals could provide more detailed
information on the electronic structure and dynamics than available
from time-resolved XANES, by probing states with multiple core
electrons excited.  Many proposed applications of the new sources make
use of their ultrashort temporal resolution and high intensity to
monitor dynamical processes in real time. Two-photon absorption
\cite{SekikawaKosugeKanaiEtAl2004} and x-ray driven molecular dynamics
\cite{GagnonRanitovicTongEtAl2007} have been demonstrated using HH
sources.  Techniques such as diffraction or pump probe do not rely on
the coherence properties of the beams. The technique considered in
this Letter, in contrast, depends also on pulse coherence in an
essential way, and should become feasible once high intensity,
attosecond
\cite{CorkumBurnettIvanov1994,BartelsBackusZeekEtAl2000,DrescherHentschelKienbergerEtAl2002,ZholentsFawley2004}
transform-limited pulses become available. Such pulses should allow to
control and manipulate the coherence of core excitations and use it as
a window into correlations between different regions of the
molecule. Similar ideas are effectively used in multidimensional NMR
spectroscopy \cite{Ernst1987} to probe correlations between spin
dynamics in controlled time periods using elaborate pulse
sequences. The signals are interpreted in terms of multiple
correlation functions which provide fundamentally new types of
information compared to one dimensional techniques. The same ideas
were recently extended to the infrared and optical regimes
\cite{Mukamel2000,AsplundZanniHochstrasser2000,TanakaMukamel2003,BrixnerStengerVaswaniEtAl2005}.

In this letter we propose a new class of two dimensional x-ray coherent
correlation spectroscopy (2DXCS) techniques and demonstrate how they could
provide a unique probe for interactions between the core-transitions and
electronic states that mediate these interactions. Infrared femtosecond 2D
techniques can excite molecular vibrations impulsively and probe the
subsequent correlated dynamics of nuclear wavepackets. Similarly attosecond
x-ray pulses resonant with core transitions can excite valence electrons
impulsively and probe correlations in dynamical events of resulting electron
wavepackets. Since core transitions are highly localized to the absorbing
atoms, these techniques also offer a high spatial resolution.

We consider a time-resolved coherent all-x-ray four-wave mixing
process carried out by subjecting the molecule to a sequence of three
pulses. The first pulse has wavevector $\boldsymbol{k}_{1} $ and
carrier frequency $\omega _{1}$, followed sequentially by the other
two pulses $\boldsymbol{k}_{2}$,$\omega _{2}$ and
$\boldsymbol{k}_{3}$, $\omega _{3}$. The signal is heterodyne-detected
with a fourth pulse $\boldsymbol{k}_{4}$,$\omega _{4}$. The coherent
nonlinear response generated in the
$\boldsymbol{k}_{I}=-\boldsymbol{k}_{1}+\boldsymbol{k}_{2}+
\boldsymbol{k}_{3}$ phase matching direction is recorded as a function
of the delays $t_{1}$, $t_{2}$, and $t_{3}$ between consecutive
pulses.  2DXCS is obtained by a Fourier transform of the signal with
respect to two delays, and displaying it as a two-dimensional
frequency correlation plot
\begin{equation*}
S_{I}^{(3)}(\Omega _{3},t_{2},\Omega _{1})\equiv \!\iint_{0}^{\infty }\!
\mathrm{d}t_{1}\mathrm{d}t_{3}S_{I}^{(3)}(t_{3},t_{2},t_{1})e^{i\Omega
_{3}t_{3}}e^{i\Omega _{1}t_{1}}.
\end{equation*}

\begin{figure}[!tbh]
\begin{center}
\includegraphics[width=3in,angle=0]{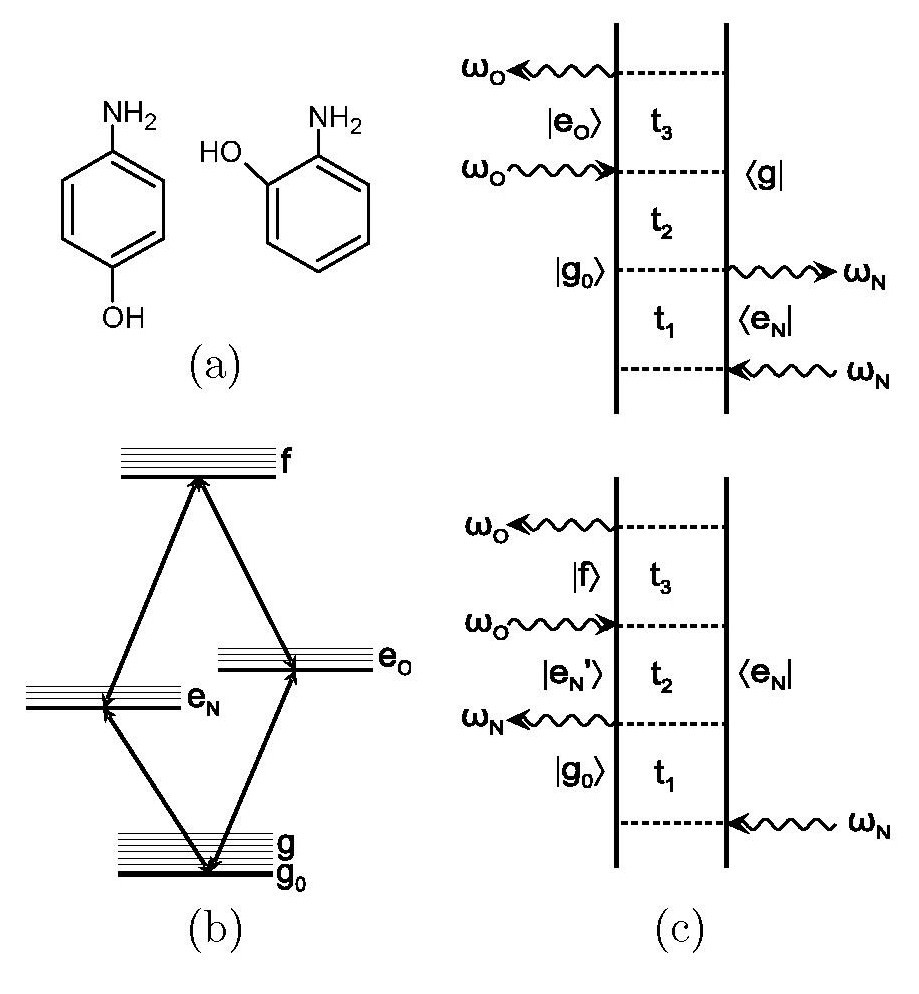}
\end{center}
\caption{(a) Para and ortho isomers of aminophenol. (b) Valence and
core-excited states of aminophenols: $g$ denotes states with no core
electron excited (including the ground state $ g_0 $), $e_N$ ($e_O$)
denotes states with the N1s (O1s) electron excited, and $f$ denotes
states with both N1s and O1s electrons excited. (c) Double-side
Feynman diagrams representing the two contributions to the cross peak
2DXCS signal [Eq.~\protect\ref{eqn:2DK1B}]}
\label{fig:fourwavemixing}
\end{figure}

Like in NMR, transitions of the same type contribute to the diagonal
part $\Omega _{1}=\Omega _{3}$ of the correlation spectrum, while
features arising from interactions among spectrally separated
transitions appear as off-diagonal cross peaks. By spreading the
signal over multiple frequency axes, the weak signatures of
interactions between different transitions can be separated from
strong same-transition signals \cite{Mukamel2000,Mukamel2005}.

We illustrate the power of these techniques for N1s and O1s
transitions of aminophenols (Figure \ref{fig:fourwavemixing}). The
polarization induced in the molecule by interactions with the x-ray
pulses is calculated in terms of nonlinear response
functions. Different time orderings appearing in the perturbative
expansion of the density matrix are described as Liouville
space-pathways and represented by double-side Feynman diagrams
\cite{Mukamel1995}. Assuming temporally well separated pulses, the
signal within the rotating-wave approximation (RWA) where we only
retain the resonant contributions is given by
\begin{eqnarray}\label{eqn:SignalK1}
  \lefteqn{
  S_{I}^{(3)} (t_3, t_2, t_1)}\\\nonumber
    &=&
        \mathrm{Im}
        R^{(3)}_{I} (t_3,t_2,t_1) e^{ i(\omega_3+\omega_2-\omega_1)t_3 + i(\omega_2-\omega_1)t_2 - i\omega_1 t_1}
\end{eqnarray}
where the response function
\begin{align}\label{eqn:ResponseK1}
 R_{I}^{(3)} &(t_3,t_2,t_1) 
      \\\nonumber
    =&
      i^3 \big[
      \average{ \hat B_1^- (0) \hat B_3^+ (t_2+t_1) \hat B_4^- (t_3+t_2+t_1) \hat B_2^+ (t_1)}
      \\\nonumber
   &+
      \average{ \hat B_1^- (0) \hat B_2^+ (t_1) \hat B_4^- (t_3+t_2+t_1) \hat B_3^+ (t_2+t_1)}
      \\\nonumber
   &-
      \average{ \hat B_1^- (0) \hat B_4^- (t_3+t_2+t_1) \hat B_3^+ (t_2+t_1) \hat B_2^+ (t_1)} \big]
\end{align}
is expressed in terms of the core exciton creation and annihilation
operators weighted by the pulse envelopes and expanded in terms of the
molecular electronic states 
\begin{equation}\label{eqn:ExcitonOperators}
  \hat B_j^{\pm} (\tau)
   \equiv
      \sum_{\mu \neq \nu}
      \eee_j (\omega_{\mu\nu} \mp \omega_j) e^{\pm i\omega_{\mu\nu} \tau - \Gamma_{\mu\nu} \tau}
      \mu_{\mu\nu} \ket{\mu} \bra{\nu}
\end{equation}
Here, $\mathcal{E}_{j}(\omega )$ is the Fourier transform of the
complex envelope of the $j$th pulse $\mathcal{E}_{j}(\omega )=\int
\mathrm{d}t\mathcal{E}_{j}(\tau -\tau _{j})e^{-i\omega (\tau -\tau
_{j})}.$ and $\omega _{j}$ its carrier frequency.

In the following simulations we assumed the pulse bandwidths (10 eV
for a 125 attosecond pulse) to be much smaller than the splitting
between the N1s and O1s transitions ($\sim $~120 eV). The carrier
frequencies $\omega _{j}$ thus select the desired type of core
transition, while the pulse envelopes $\mathcal{E}_{j}(\omega )$
control which valence excitations of the selected type contribute to
the response within the pulse bandwidths.

The diagonal part of the 2DXCS signal is obtained by tuning all four
$\omega _{j}$ either to N1s or to O1s core transitions, while the
cross peaks are obtained in the sequential two-color pulse
configuration where $ \omega _{1}=\omega _{2}\equiv \omega _{N}$ are
tuned to the N1s and $\omega _{3}=\omega _{4}\equiv \omega _{O}$ to
the O1s transition. In the latter case, the signal consists of the
ground-state bleaching (GSB) and excited-state absorption (ESA) terms
shown in Figure 1d. The former represents the effect of molecules
missing in the ground-state after two interactions with $\omega _{N}$
and cannot absorb the $\omega _{O}$ photon.  The latter represents the
additional absorption of molecules from the singly- to doubly-excited
states. The signal is given by
\begin{align}\label{eqn:2DK1B}
 &S_{I}^{(3)} (\Omega_3, t_2=0, \Omega_1)
      \nonumber\\
    =&
      \mathrm{Re} \big[
      \sum_{g, e_1} \frac{
        \eeee1{e_1 g_0} \eeee1{e_1 g} \mu_{g_0 e_1} \mu_{e_1 g} 
      }{
        \Omega_1 - \omega_1 + \omega_{e_1 g_0} + i\Gamma_{e_1 g_0}
      }
      \nonumber\\
   & \times \sum_{e_3} \frac{
        \eeee3{e_3 g} \eeee3{e_3 g_0} \mu_{g e_3} \mu_{e_3 g_0} 
      }{
        \Omega_3 + \omega_3 - \omega_{e_3 g} + i\Gamma_{e_3 g}
      }
      \nonumber\\
   &-
      \sum_{e_1, f} \frac{
        \eeee1{e_1 g_0} \eeee3{f e_1} \mu_{g_0 e_1} \mu_{e_1 f} 
      }{
        \Omega_1 - \omega_1 + \omega_{e_1 g_0} + i\Gamma_{e_1 g_0}
      }
      \nonumber\\
   &  \times \sum_{e_1^\prime} \frac{
        \eeee3{f e_1^\prime} \eeee1{e_1^\prime g_0} \mu_{f e_1^\prime} \mu_{e_1^\prime g_0}
      }{
        \Omega_3 + \omega_3 - \omega_{f e_1} + i\Gamma_{f e_1}
      }\big]
\end{align}
%
where $\mu_{ab}$ and $\Gamma_{ab}$ are the transition dipole and the
dephasing rate of the $ab$ transitions, respectively.

\begin{figure*}[!tbh]
\includegraphics[width=5in,angle=0]{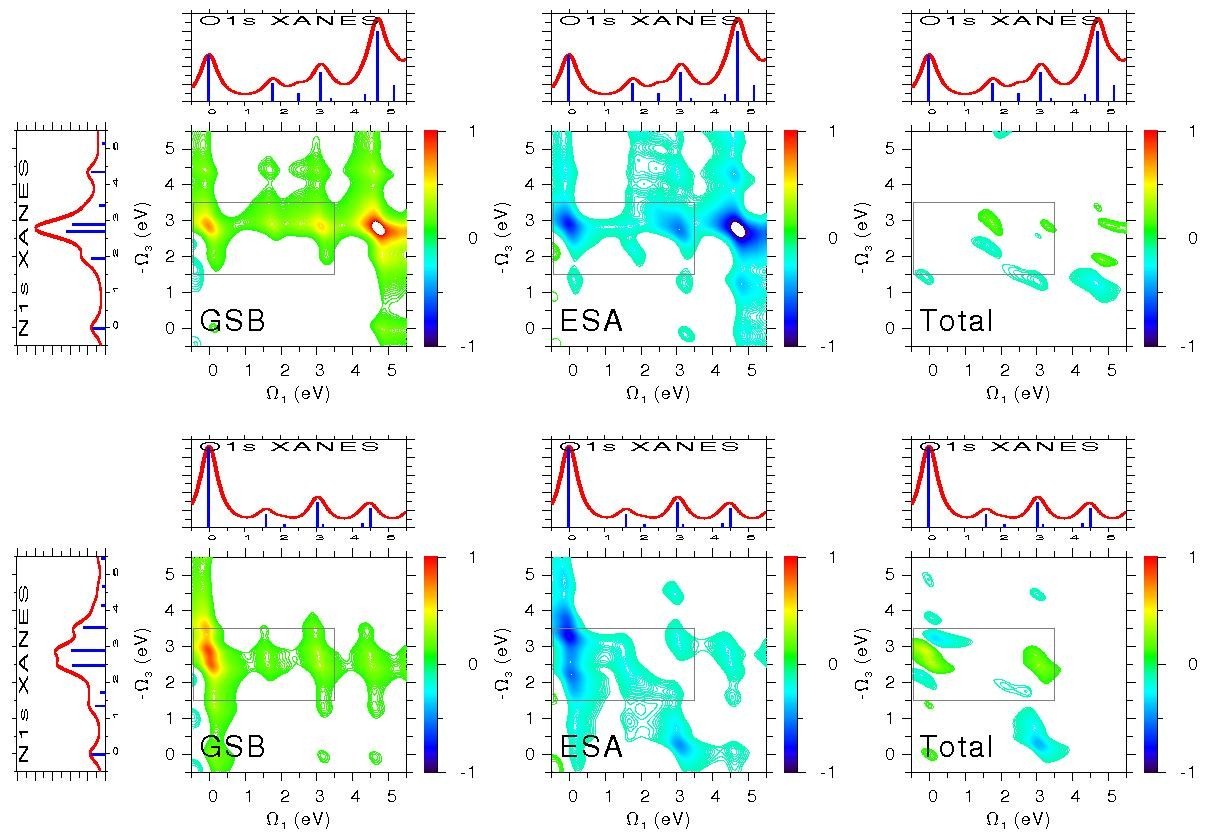}
\caption{(Color online) Simulated O1s/N1s 2DXCS cross peak ($t_2=0$)
and its GSB and ESA components for para- (upper panel) and
ortho-aminophenol (lower panel). Insets show the simulated O1s and N1s
XANES plotted as a function of $\omega-\omega_j$.}
\label{fig:crosspeak}
\end{figure*}

The cross peaks are intrinsically nonadditive and highly sensitive to
the coupling between core transitions. When the two core transitions
are decoupled, the cross peaks vanish identically due to the
interference between the one (GSB) and two-core exciton (ESA)
terms. If the core transitions are independent, we can express the
two-core exciton states as direct products
$\big|f\big\rangle=\big|e_{1}e_{3}\big\rangle$, $\omega
_{e_{3}g}=\omega _{fe_{1}}$, and the two denominators in
Eq.~(\ref{eqn:2DK1B}) become identical. Similarly the transition
dipole elements will be the same and the two terms which have an
opposite sign exactly cancel. Having distinct features in the spectra
that are induced by correlations is the main merit of 2D
techniques. In NMR, cross peaks are used to extract spin coupling and
convert this information into the structure of complex molecules.

First-principles simulations of the cross peaks dependence on the
states with two core electrons excited are particularly
challenging. We describe the core transition as a response of N+1
valence electrons to an instantly switched core hole using the
Nozieres - De Dominicus Hamiltonian
\cite{NozieresDeDominicus1969}. The core-hole potential was taken into
account by incrementing the charge of the corresponding nuclei
(equivalent-core or Z+1 approximation \cite{SchwarzBuenker1976}). The
valence excitations of the equivalent-core molecule were represented
by unoccupied Kohn-Sham orbitals. This allowed us to maintain a simple
correspondence between the molecular orbital picture and the many-body
states, which facilitated the connection between their properties and
the 2DXCS spectra.

Figure \ref{fig:crosspeak} shows the total O1s/N1s cross peak
$S_{I}^{(-3)}(\Omega _{1},t_{2}=0,\Omega _{3})$
[Eq.~(\ref{eqn:2DK1B})] and its GSB and ESA components for para and
ortho isomers of aminophenol.  The pulse carrier frequencies $\omega
_{1}$ and $\omega _{2}$ were tuned to the lowest O1s transition
($\sim$535 eV), $\omega _{3}$ and $\omega _{4}$ to the lowest N1s
transition ($\sim$401 eV). We assume the 10 eV pulse bandwidths of
($\pm$5 eV around the carrier frequencies).  The orbitals of the
original and equivalent-core molecules were calculated using the B3LYP
exchange-correlation functional and the 6-311G** set of Gaussian-type
atomic orbitals.

In both isomers, the calculated GSB component is identical to a
two-dimensional product of the O1s and N1s XANES spectra. In the para
isomer, the O and N atoms are spatially separted, and promoting the O1s
electron only weakly affects the N1s transitions. Consequently, the ESA and
GSB components are similar and the total 2DXCS cross peak is weak. In the
orho isomer, the O and N atoms are close, and exciting the O1s electron
strongly affects the N1s transitions. The ESA peaks are shifted, resulting
in a stronger 2DXCS cross peak. Thus, the 2DXCS cross peak is highly
sensitive to the relative position of the N and O atoms in the aminophenol
isomers.

\begin{figure}[!tbp]
\begin{center}
\includegraphics[width=3in,angle=0]{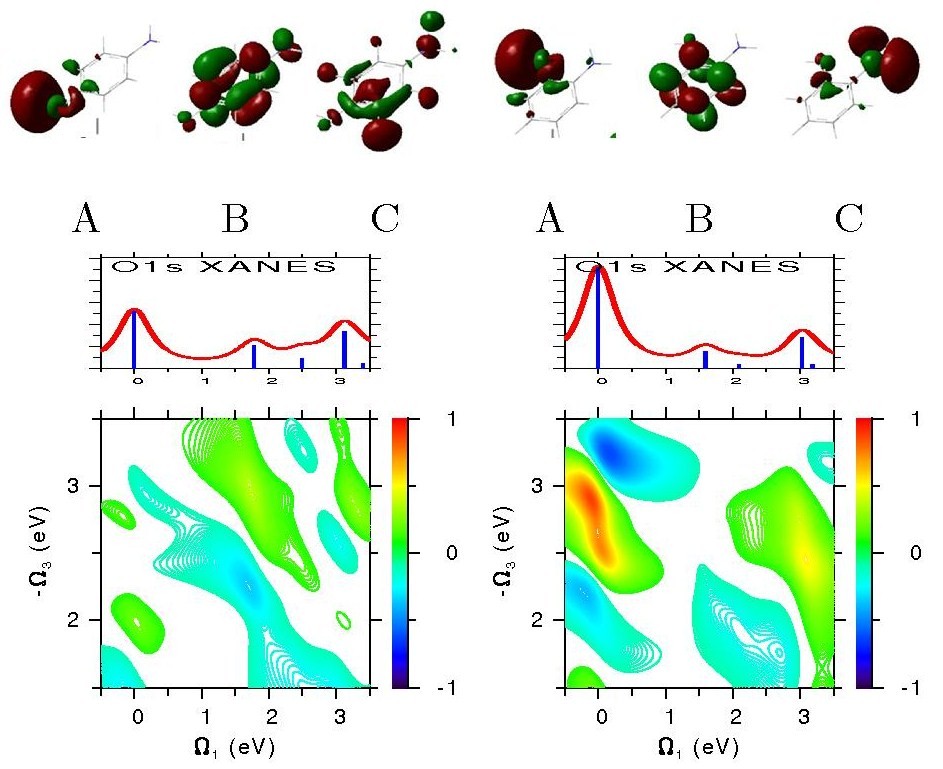}\\[0pt]
\end{center}
\caption{(Color online) Simulated O1s XANES and O1s/N1s 2DXCS cross peak (at 
$t_2=0$) of para- (left panel) and orhto-aminophenol (right panel). Insets
show the molecular orbitals populated by the promoted O1s electron for each
of core-excited states contributing to the signal.}
\label{fig:crosspeak-zoom}
\end{figure}

Figures \ref{fig:crosspeak-zoom} shows the features of the simulated
O1s XANES and O1s/N1s 2DXCS cross peak of para- and ortho-aminophenol
in the region marked on Figure \ref{fig:crosspeak}. The intensity
scale was magnified by a factor of two to better show the structure of
each peak. The equivalent-core molecular orbitals responsible for the
three stronger peaks (marked A, B, and C) in the O1s XANES are shown
as well. For example, orbital A is populated by the promoted O1s
electron in the lowest O1s transition and, in the single-orbital
approximation, represents the lowest O1s core-excited state of
para-aminophenol.

The XANES signals are insensitive to the precise form of the
wavefunctions of the corresponding states. For example, despite the
difference in the shape of orbital B between two isomers, its profile in the
vicinity of the O atom remains the same and so is the corresponding peak in
XANES. However, its contribution to the 2DXCS cross peak is markedly
different. In the para isomer, orbital B is delocalized and extends over
both O and N atoms. Correspondingly, promoting the O1s electron to this
orbital significantly affects the N1s transition, resulting in a
distinguishable cross peak. In the ortho isomer, orbital B vanishes in the
vicinity of the N atom. Promoting the O1s electron does not affect the N1s
transitions and the cross peak vanishes. The cross peak thus carries
information about the wavefunctions of these states. Similarly, despite a
stronger dipole coupling, the contribution of orbital A of para-aminophenol
to the cross peak is much weaker than that of orbital B, due to the
localization of the former to the O atom. In the ortho isomer, however, the
corresponding cross peak is strong, indicating that the atoms are
sufficiently close for the N1s transitions to be affected even though the
O1s electron remains in the vicinity of the O atom.

In conclusion, we have demonstrated how the off-diagonal part of the 2DXCS
spectra depends on the interactions between core transitions, providing
information about the electronic states that mediate this interactions.
Simulations of the cross peak 2DXCS signal of aminophenol showed that it
provides further insights into the wavefunctions of these states beyond the
dipole coupling available from XANES. The dependence of the 2DXCS spectra on
transitions of different character (valence, core transitions with one and
two core electrons excited) makes it particularly challenging for
first-principles simulations. Predicting and interpreting the 2DXCS spectra
will provide a critical test for electron structure computations and should
stimulate the development of accurate methods for excited states of
two-electron character.

The 2DXCS measurements proposed here are beyond the capabilities of
existing x-ray sources. Generating the required pulse sequences with
sub femtosecond duration, high flux and controlled timing and phase
will require to overcome many experimental difficulties. Given the
calculated values of the N 1s and O 1s transition dipole moments of
0.1 atomic units, the laser focal area of 10$^{-10}$ cm$^2$, and the
surface density on the order of 10$^{14}$ aminophenol molecules per
cm$^2$, we estimate the absorption probability to be on the order of
10$^{-5}$. In the fully resonant four-wave-mixing signal there is thus
1 generated photon per 10$^{15}$ incoming photons. Even with the XFEL
capability of generating 10$^{13}$ photons per pulse, high repetition
rates and long acquisition times will be necessary to make the
experiment feasible. Other experimental challenges include reducing
the duration of XFEL pulse \cite{ZholentsFawley2004}, mainting
longitudinal coherence, and generating multiple pulses with controlled
timing. The HH sources have the required pulse duration and are fully
spatially coherent, but have much lower brilliance
\cite{KapteynCohenChristovEtAl2007}. New phase matching technique that
works at high photon energies are being developed
\cite{CohenZhangLytleEtAl2007}. This paper aims at establishing the
theoretical basis only, and as it took 20 years to extend this NMR
radiowave technology to the infrared and the visible regions, similar
time and breakthroughs will be needed to realize these measurements in
the x-ray regime.

We wish to thank Professors P.M. Rentzepis and M. Murnane for most useful
discussions.  The support from Chemical Sciences, Geosciences and
Biosciences Division, Office of Basic Energy Sciences, Office of Science,
U.S. Department of Energy is gratefully acknowledged.

\end{document}